\documentclass[runningheads]{llncs}
\usepackage[T1]{fontenc}
\usepackage{graphicx}
\usepackage{multirow}
\usepackage{booktabs}
\usepackage{adjustbox}
\usepackage[mathscr]{euscript}
\DeclareSymbolFont{rsfs}{U}{rsfs}{m}{n}
\DeclareSymbolFontAlphabet{\mathscrsfs}{rsfs}
\usepackage{csquotes}
\usepackage{algorithmic}
\usepackage{csquotes}
\usepackage{cite}
\usepackage{amsmath,amssymb,amsfonts}
\usepackage{algorithmic}
\usepackage{graphicx}
\usepackage{textcomp}

\usepackage{bbold}
\usepackage{caption}

\usepackage{float}
\usepackage{graphicx}
\usepackage{caption}
\usepackage{array}
 \usepackage{subcaption}
 \usepackage{comment}
 \usepackage{rotating}
\usepackage{pgf}

\begin{document}
%
\title{Medi-CAT: Contrastive Adversarial Training for Medical Image Classification}
\titlerunning{Contrastive Adversarial Training for Medical Image Classification}
%
%

\author{Pervaiz Iqbal Khan\inst{1,2}\orcidID{0000-0002-1805-335X} \and
Andreas Dengel\inst{1,2}\orcidID{0000-0002-6100-8255} \and
Sheraz Ahmed\inst{1}\orcidID{0000-0002-4239-6520}}
\authorrunning{Pervaiz Iqbal Khan et al.}
%
\institute{
German Research Center for Artificial Intelligence (DFKI), Kaiserslautern, Germany \and
RPTU Kaiserslautern-Landau  \\
\email{\{pervaiz.khan,andreas.dengel,sheraz.ahmed\}@dfki.de}\
\email{}
\
}

\maketitle              
\begin{abstract}
There are not many large medical image datasets available. For these datasets, too small deep learning models can't learn useful features, so they don't work well due to underfitting, and too big models tend to overfit the limited data. As a result, there is a compromise between the two issues. This paper proposes a training strategy \textit{Medi-CAT} to overcome the underfitting and overfitting phenomena in medical imaging datasets. Specifically, the proposed training methodology employs large pre-trained vision transformers to overcome underfitting and adversarial and contrastive learning techniques to prevent overfitting. The proposed method is trained and evaluated on four medical image classification datasets from the MedMNIST collection. Our experimental results indicate that the proposed approach improves the accuracy up to $2\%$ on three benchmark datasets compared to well-known approaches, whereas it increases the performance up to $4.1\%$ over the baseline methods.

\end{abstract}

\section{\uppercase{Introduction}}
\label{sec:introduction}

The classification of medical images aids healthcare professionals in evaluating the images in a quick and error-free manner. It uses the discriminative features present in the images to distinguish between different images. Traditionally, convolutional neural networks (CNNs) have been employed to learn the image features and hence improve computer-aided diagnosis systems \cite{lo2022computer,hu2022gashissdb,hu2022application,yang2021computer}. CNNs learn the discriminative features from the images to perform tasks such as classification, object detection, etc. 

However, CNNs learn discriminative features by exploiting local image structure, and they cannot capture long-range dependencies present within the image. Recently, the Transformer methods \cite{vaswani2017attention,devlin2018bert,yang2019xlnet,radford2018improving} have revolutionized the field of natural language processing (NLP) by employing a self-attention mechanism to capture global dependencies present in the text. The success in NLP tasks has led to the suggestion of a transformer architecture for vision tasks \cite{dosovitskiy2020image}. The Vision Transformer (ViT) \cite{dosovitskiy2020image} converts an image into $16 \times 16$ patches ( like tokens in NLP tasks), and takes them as input to generate its feature representation. It has shown superior performance over the CNNs in various studies \cite{wang2021pyramid,dosovitskiy2020image}.

Large models like ViT may be prone to overfitting the smaller datasets by retaining the training examples and may fail to perform well when faced with unknown information. This can be particularly problematic in the medical imaging field, where data is scarce. Despite the large number of training samples in some datasets \cite{yang2023medmnist}, the per-class samples are still small due to the large number of classes.

In this paper, we propose a training methodology to overcome the overfitting issue by utilizing adversarial training and contrastive learning. We primarily use the Fast Gradient Sign Method (FGSM) \cite{goodfellow2014explaining} to generate adversarial examples. Then we jointly train the clean and adversarial examples to learn their representations. In addition, we use a contrastive learning method \cite{zbontar2021barlow} that improves image representation by bringing the clean and adversarial example pairs closer and pushing the other examples away from them.

The main contributions of this paper are:
\begin{itemize}
    \item It proposes a novel method for avoiding overfitting by jointly minimizing the training objective for the clean and adversarial examples.
    \item It performs experimentation on four public datasets in the domain of medical image classification to evaluate the effectiveness of our proposed training method.
    \item The proposed approach exceeds the well-known approaches in the literature on three out of four datasets.
\end{itemize}

\section{Related Work}
In this section, we review related work in the image processing domain generally and medical image classification in particular.
\subsection{Convolutional Neural Networks}
Convolutional neural networks (CNNs) have made great progress in the domain of computer vision due to their ability to learn useful image feature representation. GoogLenet \cite{szegedy2015going} used the inception network to improve feature learning. \cite{he2016deep} proposed ResNet that used residual connections to overcome the vanishing gradient problem. In order to enhance the CNNs efficiency, \cite{howard2017mobilenets} proposed MobileNet that employed both depth-wise separable convolutions and point-wise convolutions. DensNet \cite{huang2017densely} used skip-connections between every two successive layers and concatenated their features instead of their summation. ConvNext \cite{liu2022convnet} applied 7x7 depthwise convolutions and achieved comparable performance to Vision Transformers (ViT)\cite{dosovitskiy2020image}.   
\subsection{Vision Transfomers}
After achieving significant success in natural language processing (NLP), transformers in the image domain, i.e., vision transformers (ViT) have been successfully implemented in various tasks, including image classification \cite{dosovitskiy2020image}, image segmentation \cite{zheng2021rethinking}, and object detection \cite{carion2020end}.  ViT divides an image into patches, which resemble tokens in NLP, and then applies transformer layers to uncover the correlation between these patches. This way, it learns useful features for the downstream tasks. Many improvements have been proposed over the standard ViT. To strengthen the local structural relationship between the patches, T2T-ViT \cite{yuan2021tokens} generates tokens and then combines neighboring tokens into a single token. Swin Transformer \cite{liu2021swin} learns the in-window and cross-window relationships by applying self-attention in the local window with the shifted window. The pooling-based vision transformer (PiT) \cite{heo2021rethinking} uses a newly designed pooling layer in the transformer architecture to reduce spatial size similar to CNNs and empirically shows the improvement.     

\subsection{Medical Image Classification}
 \cite{yang2023medmnist} introduced the MedMNIST dataset collection, which comprises 12 datasets related to 2D images and 6 datasets related to 3D images. They performed classification tasks on these datasets using various models such as ResNet-18 \cite{he2016deep}, ResNet-50 \cite{he2016deep}, auto-sklearn \cite{feurer2015efficient}, AutoKeras \cite{jin2019auto}, and Google AutoML Vision \cite{bisong2019building}. MedViT \cite{manzari2023medvit} proposed a hybrid model that combines the capabilities of CNNs to model local representations with the capabilities of transformers to model the global relationship. Their attention mechanisms use efficient convolution to solve the problem of quadratic complexity. A novel mixer, known as a C-Mixer, has been proposed by \cite{zheng2023complex} that incorporates a pre-training mechanism to address the uncertainty and inefficient information problem in label space. This mixer employs an incentive imaginary matrix and a self-supervised method with random masking to overcome the uncertainty and inefficient information problem in label space. BioMedGPT \cite{zhang2023biomedgpt}, is a generalized framework for multi-modal tasks in the medical domain, such as images and clinical notes. It first employs pre-training using masked language molding (MLM),  masked image infilling, question answering, image captioning, and object detection to learn diverse types of knowledge. Then it is fine-tuned to the downstream tasks to show the efficacy of the model for transferring knowledge to other tasks.
 \section{\uppercase{Methodology}}

\begin{figure*}
\includegraphics[width=\textwidth]{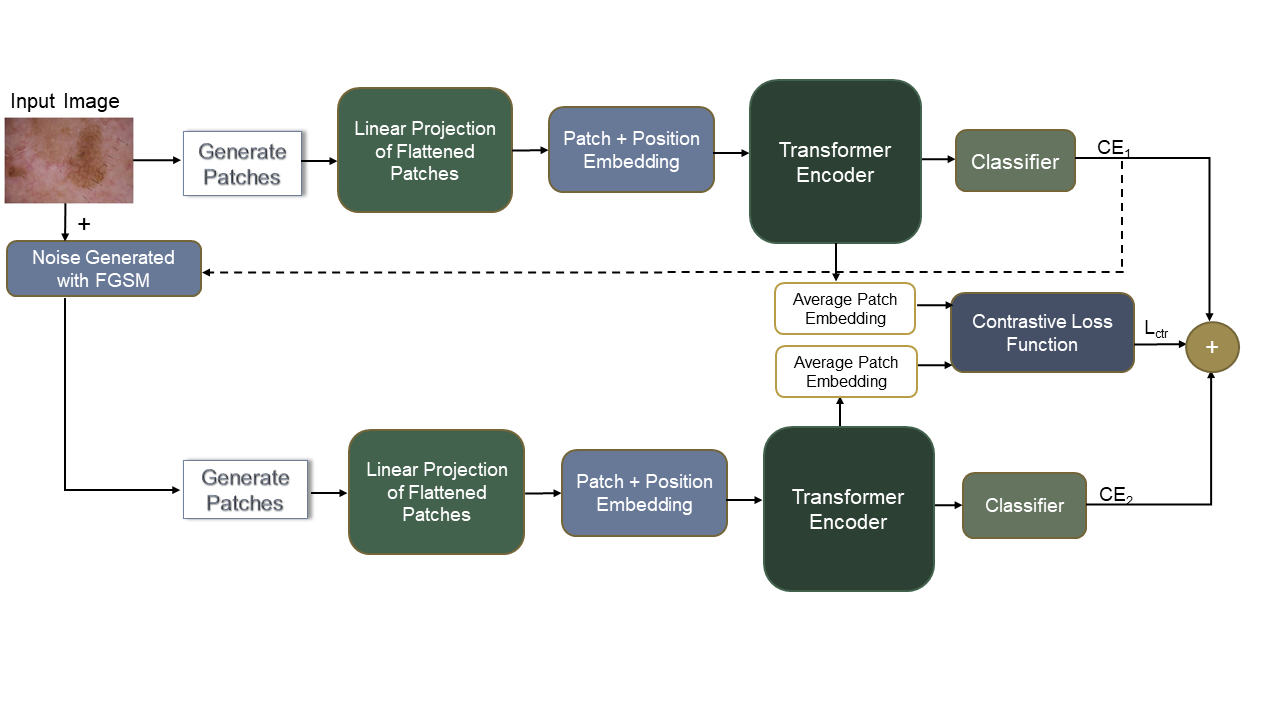}
\caption{Proposed training methodology for medical image classification.} \label{fig:method}
\end{figure*}
In this section, we present our proposed training method for medical image classification. As shown in Figure \ref{fig:method}, our method consists of three main components. (1) Transformer-based image encoder that extracts features from the input image; (2) image encoder that takes images with perturbations generated by Fast Gradient Sign Method (FGSM) \cite{goodfellow2014explaining} method and extracts features; (3) Contrastive loss that takes the average patch embeddings of the clean and perturbed images as input and further improves their features in the representation space.
\subsection{Image Enocder}\label{encoder}
The pre-trained Vision Transformer (ViT) \cite{dosovitskiy2020image} is chosen as the image encoder to encode the image in the representation space. An image is first split into $16 \times 16$ patches as tokens, and then those tokens are passed as inputs to the ViT. At the end of its forward pass, ViT returns the classification loss computed using cross-entropy as given by the following equation:  

 \begin{equation}\label{eq:loss}
\mathcal{L}_{CE} = - {\dfrac{1}{N}}\sum_{i=1}^{N}\sum_{c=1}^{C}y_{i,c}log(p(y_i,c|s^{i}_{[CLS]}))
    \end{equation}
    where $s^{i}_{[CLS]}$ is the final hidden representation for the  $i$-th training example in the batch, `N' is the number of training examples in the batch, and `C' is the number of classes.

\subsection{Adversarial Examples}
\label{sec:perturb}
Adversarial examples are generated by adding a small amount of perturbations in the images from the training set. We utilize the FGSM \cite{goodfellow2014explaining} method to generate the amount of noise $\eta$. Let $f_{\theta}(x_i, y_i)$ be a neural network parameterized by $\theta$ where $x_i$, and $y_i$ represent the input example and its corresponding label, respectively. Let, $\mathcal{L}$ represent the loss at the end of the forward pass as calculated using equation \ref{eq:loss}. Then perturbation $\eta$ generated by FGSM \cite{goodfellow2014explaining} is given as follows:
\begin{equation}\label{eq:fgsm}
    \eta = -\epsilon sign(\nabla_{x_i}\mathcal{L}(f_\theta (x_i), y_i))
\end{equation}
In equation \ref{eq:fgsm}, $\nabla$ is the gradient of the loss $\mathcal{L}$ w.r.t input $x_i$. $\epsilon$ is the hyperparameter controlling the amount of noise. The generated noise $\eta$ is added to the input image to generate an adversarial example. The generated adversarial example is passed to the image encoder as discussed in section \ref{encoder}, where another forward pass is completed and another classification loss is computed as given by equation \ref{eq:loss}. We use the shared image encoder to extract the representations for the clean and perturbed images.

\subsection{Contrastive Learning}
We employed Barlow Twins \cite{zbontar2021barlow} as a contrastive learning method that takes two inputs, i.e., encoding of the clean image, and encoding of its perturbed version that are generated by image encoder. The encoding of the last hidden state of the image encoder can be represented as $H \in \mathbb{R}^{p \times d}$. Here, $p$ is the number of patches, i.e. 16, and $d$ is the number hidden units of ViT, i.e., $1024$. We average the encoding of all the patches for both clean and perturbed examples and then pass it to Barlow Twins \cite{zbontar2021barlow} loss function that improves their representations by pulling the pair of clean and perturbed encoding closer while pushing them away from other image encodings in the training batch.

Let $E^{o}$ and $E^{p}$ represent the averaged encoding of the original and its perturbed version, respectively.  Then, the Barlow Twins \cite{zbontar2021barlow} improves their representations by using following objective function:
\begin{equation}
 \mathcal{L_{CTR}} = \sum_{i=1} (1 - X_{ii})^2 + \lambda \sum_{i=1} \sum_{j \neq i}X_{ij}^2
\end{equation}
where $\sum_{i=1} (1 - X_{ii})^2$, and $\sum_{i=1} \sum_{j \neq i}X_{ij}^2$ are the invariance, and redundancy reduction terms respectively, and `$\lambda$' controls weights between the two terms. The matrix `X' computes the cross-correlation between $E^{o}$, and $E^{p}$. The matrix `X' is given as follows:

\begin{equation}
    X_{ij} = \frac{\sum_{b=1}^{N} E^{o}_{ b, i} E_{b, i}^{p} }{\sqrt{ \sum_{b=1}^{N} (E^{o}_{b, i} })^2    \sqrt{ \sum_{b=1}^{N} (E^{p}_{b, i} })^2 }
\end{equation}
where `b' is the batch size, and `$X_{ij}$' represents the entry of the i-th row and j-th column of the correlation matrix`X'. Both $E^{o}$ and $E^{p}$ $\in  \mathbb{R}^{1 \times 1024}$
\subsection{Training Objective}

The training objective of our proposed method consists of three parts:(1) Minimizing the classification loss of the clean images; (2) minimizing the classification loss of the perturbed images; (3) minimizing the contrastive loss for the clean and perturbed image encodings.

Total loss $\mathcal{L}$ is given as follows:
\begin{equation}
\mathcal{L} = \frac{(1- \alpha )}{2} (\mathcal{L}_{CE_1} + \mathcal{L}_{CE_2}) + \alpha \mathcal{L}_{CTR}
\end{equation}
where `$\mathcal{L}_{CTR}$' is the contrastive loss \cite{zbontar2021barlow}, `$\mathcal{L}_{CE_1}$', and `$\mathcal{L}_{CE_2}$' are two classification losses for the clean and perturbed images, and `$\alpha$' is the trade-off parameter between the three losses. A higher value of `$\alpha$' means more weight to the contrastive loss.

\section{\uppercase{Experiments}}
In this section, we present the datasets, evaluation metrics, and training details that are used for the experimentation.

\subsection{Datasets}\label{ds}
MedMNIST \cite{yang2023medmnist} is a collection of 2D and 3D medical images related to ordinal regression, multi-label, and multi-class classification. We performed experimentation on four multi-class classification datasets from the MedMNIST \cite{yang2023medmnist} dataset to validate the performance of our proposed training strategy. The details of each dataset are given in Table \ref{tbl:ds}.
\begingroup
\setlength{\tabcolsep}{6pt} 
\renewcommand{\arraystretch}{1.4}
\begin{table*}[]
\centering
\caption{Statistics of datasets from MedMNIST \cite{yang2023medmnist} collection used in our experiments .}
\label{tbl:ds}
\begin{tabular}{lllll}
\toprule
\hline
Name         & Modality & $\#$ Classes &  $\#$ Samples & Train/validation/Test \\ \hline
DermaMNIST \cite{yang2023medmnist} & Dermatoscope             & 7 &  10,015 & 7,007/1,003/2,005             \\ \hline
OrganAMNIST \cite{yang2023medmnist} & Abdominal CT             & 11    & 58,850 & 34,581/6,491/17,778           \\ \hline

OrganCMNIST \cite{yang2023medmnist}    & Abdominal CT             & 11  & 23,660       &  13,000/2,392/8,268      \\ \hline
OrganSMNIST \cite{yang2023medmnist}     & Abdominal CT             & 11 &  25,221   &  13,940/2,452/8,829            \\ \hline
\bottomrule
\end{tabular}
\end{table*}
\endgroup
\subsection{Evaluation Metrics}
Following \cite{zhang2023biomedgpt}, we use accuracy as an evaluation metric. Accuracy is based on the threshold used to evaluate the discrete label prediction and is sensitive to class imbalance.  For the datasets we used in our experiments, there is no class imbalance, so accuracy is a good metric. On each dataset, we report the average accuracy score for two random runs with seeds of $42$, and $44$ respectively.

\subsection{Training Details}
We conducted training on each of the datasets mentioned in the section \ref{ds} for $50$ epochs, with a batch size of $48$. Before the training, all images were resized to 224x224 pixels. We used the same parameters as in \cite{yang2023medmnist} to normalize all the images. We used a fixed learning rate of $1e^{-4}$ and AdamW \cite{loshchilov2018fixing} as an optimizer in all our experiments. The cross-entropy and the Barlow Twins \cite{zbontar2021barlow} were employed as classification loss and contrastive loss, respectively. The default hyperparameters were used for contrastive loss, and unlike the original implementation, we did not use a projection network for its two inputs. We performed a grid search for $\alpha \in \{ 0.1, 0.2, 0.3, 0.4, 0.5, 0.6, 0.7, 0.8, 0.9 \}$ and $\epsilon \in \{ 0.0001, 0.001, 0.0005, 0.001 \}$ and used the validation set model with the highest accuracy for test set evaluation.

\section{\uppercase{Results and Analysis}}
In this section, we present the results and analysis of our proposed approach. Furthermore, we compare our results with well-known approaches in the literature and also discuss the effect of various hyper-parameters on the model performance.
\subsection{Comparsion with Existing Methods}
Table \ref{tbl:main-results} shows that our proposed method outperforms the existing methods on three datasets, whereas it remains second-best on the fourth one. These enhancements can be attributed to adversarial training and contrastive learning, which enhance the generalization of the model by avoiding overfitting. However, these improvements come with additional training costs, which are incurred by gradient calculations in FGSM \cite{goodfellow2014explaining} method and additional training passes with perturbed images. However, accuracy can be more important in health-related tasks than training costs.
\begingroup
\setlength{\tabcolsep}{4pt} 
\renewcommand{\arraystretch}{1.6}
\begin{table*}[]
\centering
\caption{compares the results of our proposed method with existing methods in literature on DermaMNIST, OrganAMNIST, OrganCMNIST, and OrganSMNIST \cite{yang2023medmnist} datasets in terms of accuracy score. Similar to \cite{zhang2023biomedgpt}, we only present SotA approaches if they provided open-source code for reproducibility. The proposed method outperforms existing methods on three out of four datasets.}
\label{tbl:main-results}
\begin{tabular}{lllllllll}
\toprule
\hline
\multirow{1}{*}{Methods} & \multicolumn{1}{l}{DermaMNIST} & \multicolumn{1}{l}{OrganAMNIST} & \multicolumn{1}{l}{OrganCMNIST} & \multicolumn{2}{l}{OrganSMNIST} \\
\toprule
                         \hline
ResNet-18 (28) \cite{yang2023medmnist}              & 0.735                   & 0.935                       & 0.900               & 0.782              \\ \hline
ResNet-18 (224) \cite{yang2023medmnist}             & 0.754                   & 0.951                      & 0.920          & 0.778                 \\ \hline
ResNet-50 (28)  \cite{yang2023medmnist}              & 0.735                   & 0.935                        & 0.905          & 0.770                \\ \hline
ResNet-50 (224) \cite{yang2023medmnist}            & 0.731                  & 0.947                   & 0.911             & 0.785             \\ \hline
auto-sklearn  \cite{yang2023medmnist}                & 0.719                    & 0.762                     & 0.829              & 0.672              \\ \hline
AutoKeras   \cite{yang2023medmnist}                 & 0.749                    & 0.905                    & 0.879               & 0.813             \\ \hline
Google AutoML Vision  \cite{yang2023medmnist}       & 0.768                   & 0.886                       & 0.877               & 0.749              \\ \hline
FPVT \cite{liu2022feature}                         & 0.766                   & 0.935                       & 0.903              & 0.785             \\ \hline
MedVIT-T (224) \cite{manzari2023medvit}              & 0.768             & 0.931                      & 0.901              & 0.789              \\ \hline
MedVIT-S (224)  \cite{manzari2023medvit}            & 0.780                   & 0.928                      & 0.916               & 0.805             \\ \hline
MedVIT-L (224)  \cite{manzari2023medvit}            & 0.773                    & 0.943                      & 0.922                & 0.806              \\ \hline
Complex Mixer \cite{zheng2023complex}            & \textbf{0.833}           & 0.951                      & 0.922          & 0.810              \\ \hline
BioMed-GPT \cite{zhang2023biomedgpt}                     & 0.786                       & 0.952                           & 0.931                   & 0.823              \\ \hline
Ours                         & 0.824                    & \textbf{0.961}                  & \textbf{0.940}                    & 
\textbf{0.843}    
\\
\hline
\bottomrule
\end{tabular}
\end{table*}
\endgroup

\subsection{Analysis of Noise Amount and Trade-off Parameter}
\begin{figure*}[]
        \centering
        
         \begin{subfigure}[t]{0.99\textwidth}
            \begin{subfigure}[t]{0.49\textwidth}
                \includegraphics[width=\textwidth, height=4cm, keepaspectratio]{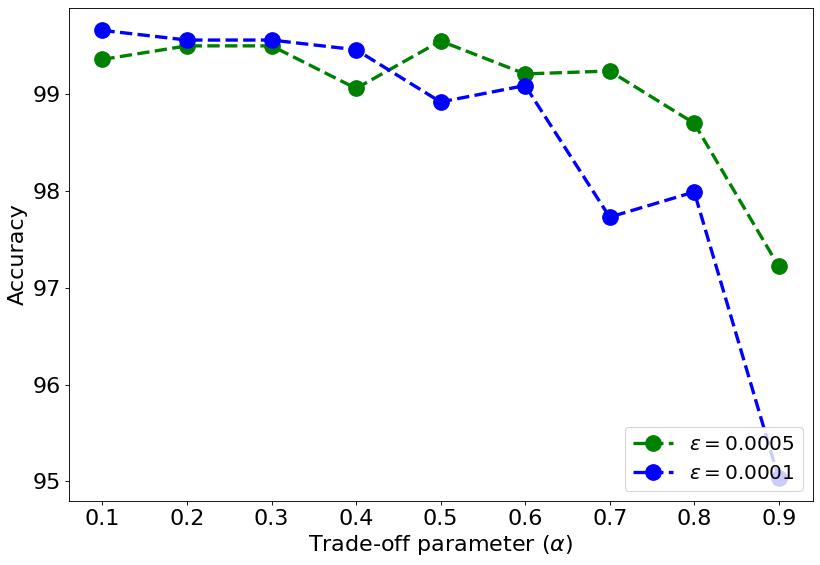}
                \caption{ Plots showing the effect of trade-off parameter $\alpha$ and noise controlling parameter $\epsilon$ on organamnist dataset.}
                \label{cap:plots1}
            \end{subfigure}
           \hfill
            \begin{subfigure}[t]{0.49\textwidth}
                \includegraphics[width=\textwidth , height=4cm, keepaspectratio]{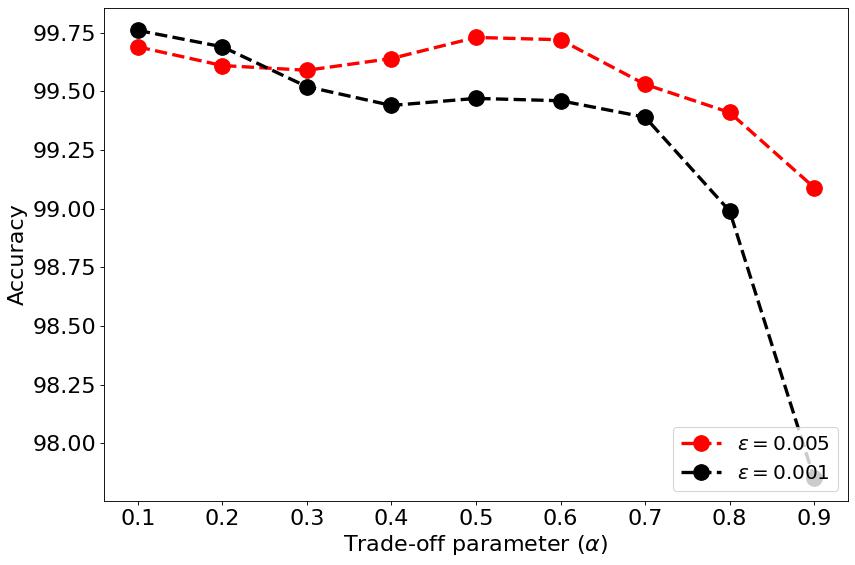}
                \caption{ Plots showing the effect of trade-off parameter $\alpha$ and noise controlling parameter $\epsilon$ on organamnist dataset.}
                \label{cap:plots2}
            \end{subfigure}
        \end{subfigure}

     \begin{subfigure}[t]{0.99\textwidth}
            \begin{subfigure}[t]{0.49\textwidth}
                \includegraphics[width=\textwidth, height=4cm, keepaspectratio]{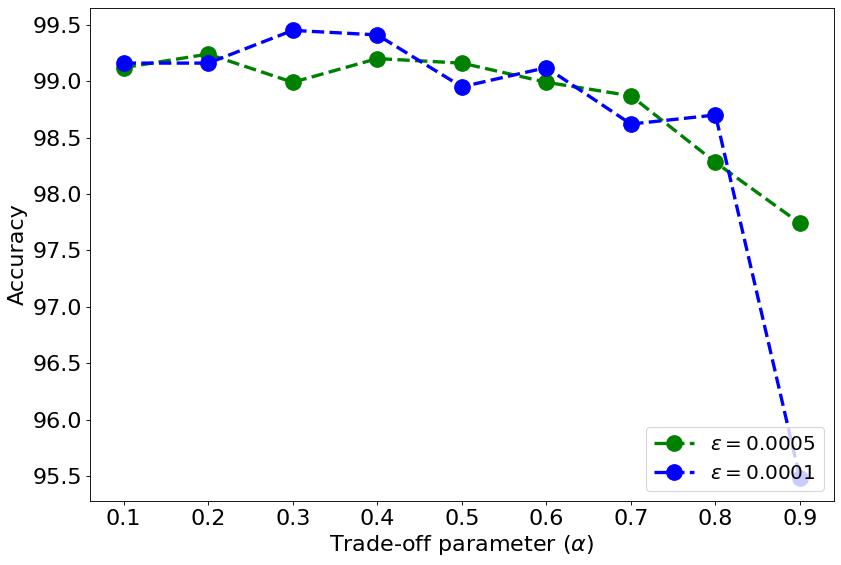}
                \caption{ Plots showing the effect of trade-off parameter $\alpha$ and noise controlling parameter $\epsilon$ on organcmnist dataset.}
            \end{subfigure}
           \hfill
            \begin{subfigure}[t]{0.49\textwidth}
                \includegraphics[width=\textwidth , height=4cm, keepaspectratio]{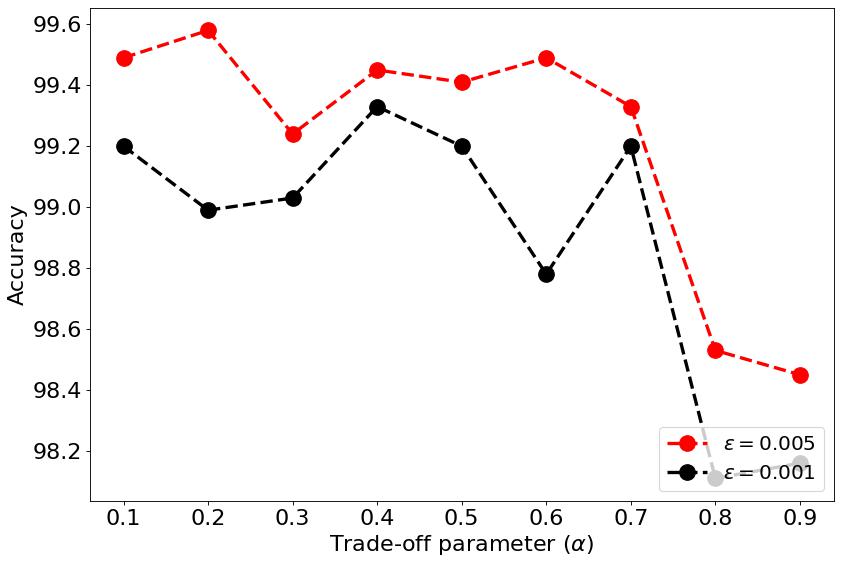}
                \caption{ Plots showing the effect of trade-off parameter $\alpha$ andnoise controlling parameter $\epsilon$ on organcmnist dataset.}
            \end{subfigure}
        \end{subfigure}
 \begin{subfigure}[t]{0.99\textwidth}
            \begin{subfigure}[t]{0.49\textwidth}
                \includegraphics[width=\textwidth, height=4cm, keepaspectratio]{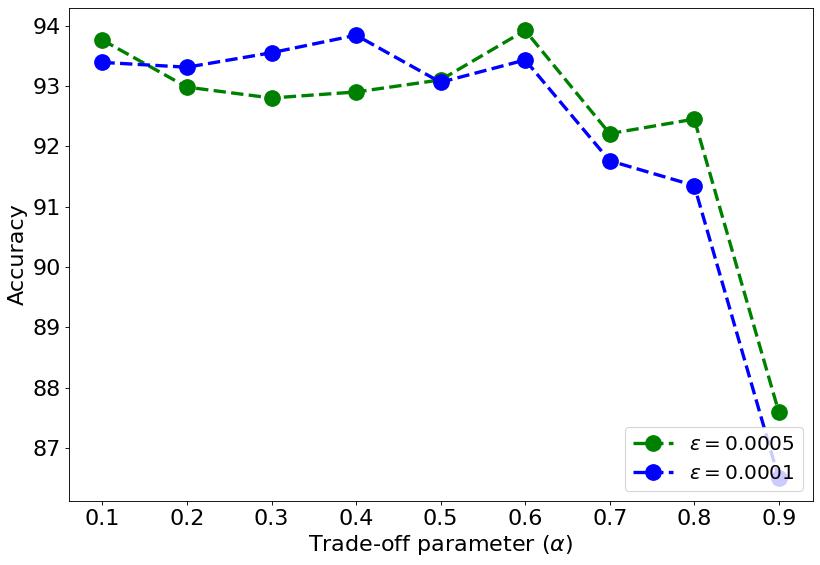}
                \caption{ Plots showing the effect of trade-off parameter $\alpha$ and noise controlling parameter $\epsilon$ on organsmnist dataset.}
            \end{subfigure}
           \hfill
            \begin{subfigure}[t]{0.49\textwidth}
                \includegraphics[width=\textwidth , height=4cm, keepaspectratio]{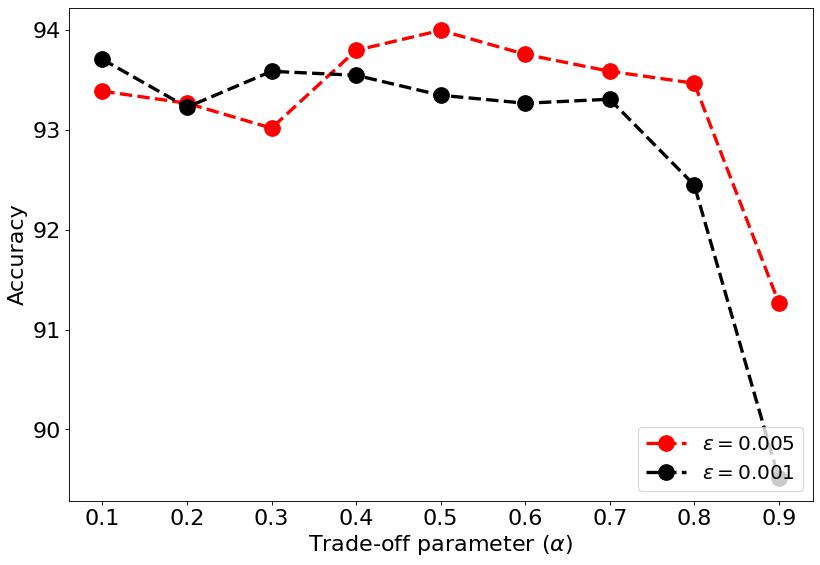}
                \caption{ Plots showing the effect of trade-off parameter $\alpha$ and noise controlling parameter $\epsilon$ on organsmnist dataset.}
            \end{subfigure}
        \end{subfigure}
\begin{subfigure}[t]{0.99\textwidth}
            \begin{subfigure}[t]{0.49\textwidth}
                \includegraphics[width=\textwidth, height=4cm, keepaspectratio]{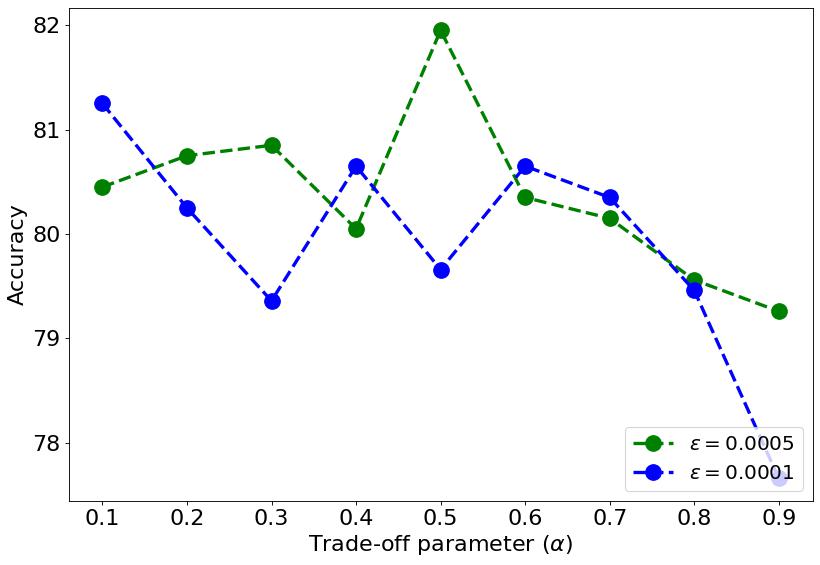}
                \caption{ Plots showing the effect of trade-off parameter $\alpha$ and noise controlling parameter $\epsilon$ on dermamnist dataset.}
                \label{cap:plots7}
            \end{subfigure}
           \hfill
            \begin{subfigure}[t]{0.49\textwidth}
                \includegraphics[width=\textwidth , height=4cm, keepaspectratio]{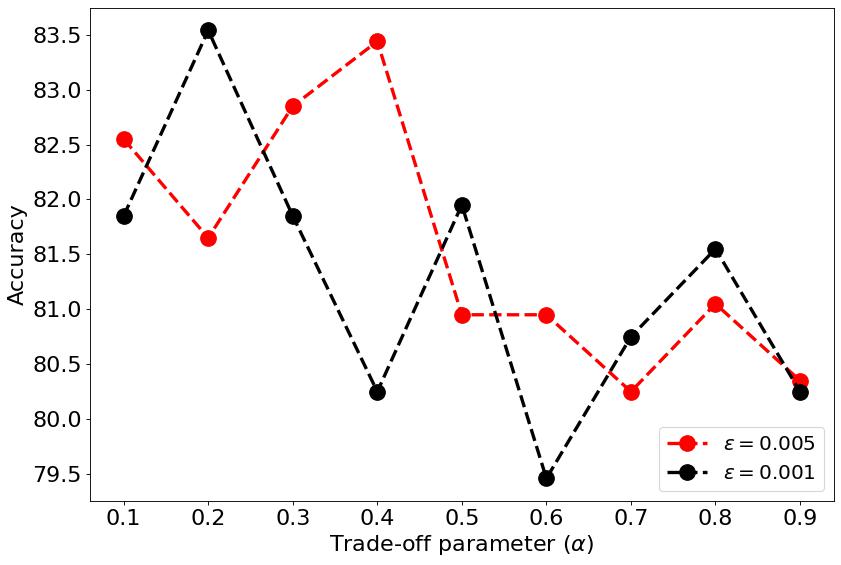}
                \caption{ Plots showing the effect of trade-off parameter $\alpha$ and noise controlling parameter $\epsilon$ on organsmnist dataset.}
                \label{cap:plots8}
            \end{subfigure}
            
        \end{subfigure}
\caption{Accuracy plots on the validation set for Medmnist datasets showing the effect of trade-off parameter and noise amount.}
        \label{cap:plots}
    \end{figure*}

Figure \ref{cap:plots} shows the effect of trade-off parameter $\alpha$ and noise controlling parameter $\epsilon$ on the validation sets of four datasets. For simplicity, these results are taken from one of the training runs. All the plots show that the accuracy for the smaller values of $\alpha$ is generally higher, whereas it decreases sharply for $\alpha \geq 0.6$. This implies giving more weight to contrastive loss after a certain degree negatively affects performance. For values of $\alpha < 0.6$ there is only a slight change in the performance of the model. As shown in Figure \ref{cap:plots7} and \ref{cap:plots8}, DemraMNIST \cite{liu2022feature} is more sensitive to both $\alpha$ and $\epsilon$ values as compared to other datasets.

\begingroup
\setlength{\tabcolsep}{4pt} 
\renewcommand{\arraystretch}{1.6}
\begin{table*}[]
\centering
\caption{shows accuracy scores on four datasets for the proposed method. The baseline model is ViT\textsubscript{Large} \cite{dosovitskiy2020image}. AT stands for Adversarial Training.}
\label{tbl:ablation-results}
\begin{tabular}{lllll}
\toprule
\hline
Method                       & DermaMNIST & OrganAMNIST & OrganCMNIST & OrganSMNIST \\
\toprule
\hline
ViT\textsubscript{Large} \cite{dosovitskiy2020image} (Baseline)                    & 0.783      & 0.954       & 0.937       & 0.841       \\ \hline
AT Only                     & 0.817      & 0.949       & \textbf{0.942}       & 0.841       \\ \hline
AT + Contrastive (Proposed) & \textbf{0.824}      & \textbf{0.961}       & 0.940       & \textbf{0.843}      \\
\hline
\bottomrule
\end{tabular}
\end{table*}
\endgroup

\subsection{Effectiveness of Proposed Method}
Table \ref{tbl:ablation-results} illustrates the effectiveness of our proposed method. The results show that incorporating adversarial training enhances the model's precision on the DermaMNIST \cite{yang2023medmnist} dataset. Furthermore, incorporating contrastive learning further improves the performance of the model. This performance enhancement of over $4\%$ can be attributed to adversarial and contrastive training. Since the original dataset size is smaller as compared to other datasets, the FGSM \cite{goodfellow2014explaining} method generates new training samples with small perturbations, and then adversarial training and contrastive learning improve feature representations. For the OrganAMNIST \cite{yang2023medmnist} dataset, adversarial training results in a decrease in model performance, whereas the addition of contrastive training enhances the performance compared to the baseline model. The inclusion of contrastive learning results in a slight decrease in performance compared to adversarial training for the OrganCMNIST \cite{yang2023medmnist}  dataset. Our method for OrganSMNIST \cite{yang2023medmnist} only makes a small improvement over the standard model. The difficulty of the dataset itself might be the reason for this, as it doesn't allow noise to improve performance.

\section{\uppercase{Conclusions}}
\label{sec:conclusion}
In this paper, we proposed a training method to overcome the problems of underfitting and overfitting in medical image classification. We used the power of a vision transformer to learn the features for different classes and then fine-tuned it on the downstream classification task. To fix the overfitting problem, we added perturbations to the training images and then jointly trained both clean and perturbed images. To improve the feature representation, we added contrastive loss that pushes the clean and perturbed versions of the sample closer and farther than the other samples in the representation space. Extensive experiments on the four benchmark medical image classification datasets demonstrate the effectiveness of our proposed method. In the future, we intend to apply the proposed method to object detection and segmentation tasks.

%
%
%
%

\bibliographystyle{splncs04}
\bibliography{references}
\end{document}